\documentclass[aps,prl,reprint,amsmath,amssymb,graphicx,longbibliography]{revtex4-1}
\usepackage{physics}
\usepackage{bm}
\usepackage{amsmath}
\usepackage{textcomp}
\usepackage{tcolorbox}
\usepackage{tikz}
\usepackage{emf}

\begin{document}

\title{Comment on Noise-Induced Subdiffusion in Strongly Localized Quantum Systems}

\author{Zahra Mohammaddoust Lashkami}%
 \email{z.mohammaddoust@znu.ac.ir}
\affiliation{%
University of Zanjan
}%

\author{Younes Younesizadeh}
\email{younesizadeh88@gmail.com}
 \affiliation{Isfahan University of Technology}

\author{Ehsan Gholami}
 \thanks{corresponding author}%
 \email{egholami.b@gmail.com}
 \affiliation{Isfahan University of Technology}

\date{\today{}}

\begin{abstract}

\end{abstract}


\maketitle
Gopalakrishnan, Islam, and Knap (hereafter GIK) adopted a perturbative approach employing a method which had been presented in Ref. \cite{Amir2009} to investigate the transport in localized systems coupled to non-Markovian dephasing noise in the deeply localized limit where the single-particle hopping is the smallest energy scale \cite{original}. As they stated in their letter this analytical exploration works when the ratios of the scales $W$; $\Lambda$;  are arbitrary, so long as each is much larger than J; i.e, $J\ll \Lambda, W$ (see the section on perturbative treatment in their letter and the supplemental material). All their subsequent analytical and numerical calculations are limited to this condition (the deeply localized limit). Nevertheless, in the majority of their results, they used this approximation outside this limit which makes their results wrong. We point at these wrong results and propose the changes in the parameter values which can give correct results. Moreover, GIK need to remove some statements as they will no longer be valid with the new parameter values. \\
All the results in Fig.2 except for Fig.2(c) when $\Lambda=20J$ (black plot) are incorrect as they do not fit in the deeply localized limit. GIK need to recalculate the results in Fig.2(a) for $\Lambda\gtrsim 5J$ and $W\gtrsim 5J$ i.e, $W=4J, 6J, 8J,10J$ , the results in Fig.2(b) should be recalculated for for $\Lambda\gtrsim 5J$ and $W\gtrsim 5J$ i.e, $\Lambda=5J, W=5J$ and the results Fig.2(c) should be recalculated for $\Lambda\gtrsim 5J$ i.e, $\Lambda=5J, 10j, 20J$. Likewise, all the sub-diffusion exponent's data points in Fig.3, except the black points when $W/J>5$ do not fit in the deeply localized limit. GIK need to recalculate short-time subdiffusion exponent for $\Lambda\gtrsim 5J$ i.e, $\Lambda=5J, 10J, 20J$ in the disorder range $W/J\gtrsim 5$. In the caption for Fig.3, GIK claimed that in the weak noise limit $\Lambda \precsim J$, the exponent depends greatly on the disorder strength $W$ approaching zero with increasing $W$. However, one can not conclude anything from their calculations about the exponents in the weak noise limit as it is outside the deeply localized limit. The exponents rest constant for the changes in the noise strength within the deeply localized limit. Regarding the results for the asymptotic diffusion constant, the results fit in the deeply localized limit, only for $W=4J$ and $\Lambda/J\gtrsim 5$. GIK need to recalculate the diffusion constants for $W\gtrsim 5J$ i.e, $W=4J, 6J, 8J,10J$ and in the noise strength range $\Lambda/J\gtrsim 5$. Once they do these corrections the diffusion constant change linearly with regard to the noise strength and there is no asymptotic behavior here.\\
At the end of the section on subdiffusive transport in the supplemental material, GIK expressed that $\tau$ has to be large enough to enable this anomalous transport regime. We want to point that outside the deeply localized regime when $\Lambda$ or $W$ are not much greater than $J$, even for white noise ($\tau=0$), such an anomalous regime could exist. Regarding stretched-exponential decay of the imbalance in a noisy environment (the results in Fig.S1), the plot for $\Lambda=20J$ is reliable, yet, they need to recalculate the results for $\Lambda\gtrsim 5$ i.e, $\Lambda=5J, 10J, 20J$ to fit in the deeply localized limit where their calculations are valid. The results in Fig.S2 (a) should be recalculated for $\Lambda\gtrsim 5$ in the disorder range $W/J\gtrsim 5$ and the results in Fig.S2 (b) are valid only in the disorder range $W/J\gtrsim 5$. GIK need to recalculate the results in FIG.S3(a), for $\Lambda\gtrsim 5J$ i.e, $\Lambda= 4J, 8J, 20J$. \\ 
In summary, the small hopping approximation is very useful in deeply localized systems but like all the other approximations, it should be used in cases that satisfy its conditions. We pointed that majority of GIK's results are calculated for the parameter values which lay outside the deeply localized regime. They need to correct their results by recalculating them for parameter values that fit in this regime and adopt their conclusions to the new corrected results.

\bibliography{sample-rmp-revtex4}

\end{document}